\documentclass[conference]{IEEEtran}
\IEEEoverridecommandlockouts
\makeatletter
\newcommand{\linebreakand}{%
  \end{@IEEEauthorhalign}
  \hfill\mbox{}\par
  \mbox{}\hfill\begin{@IEEEauthorhalign}
}
\makeatother
\usepackage{cite}
\usepackage{booktabs} 
\usepackage{amsmath,amssymb,amsfonts}
\usepackage{algorithmic}
\usepackage{graphicx}
\usepackage{textcomp}
\usepackage{xcolor}
\def\BibTeX{{\rm B\kern-.05em{\sc i\kern-.025em b}\kern-.08em
    T\kern-.1667em\lower.7ex\hbox{E}\kern-.125emX}}
\begin{document}

\title{Design and Validation of Learning Aware HMI For Learning-Enabled Increasingly Autonomous Systems*
\thanks{*We would like to thank Nandith Narayan for his initial support in setting up the infrastructure for LEIAS. This work was funded by award 80NSSC20M0005 to the NASA Langley Research Center from the NASA Shared Services Center.}
}

\author{

\IEEEauthorblockN{Parth Ganeriwala}
\IEEEauthorblockA{\textit{Computer Engineering and Sciences} \\
\textit{Florida Institute of Technology}\\
Melbourne FL, USA \\
pganeriwala2022@my.fit.edu}
\and

\IEEEauthorblockN{Michael Matessa}
\IEEEauthorblockA{\textit{Applied Research and Technology} \\
\textit{Collins Aerospace}\\
Cedar Rapids IA, USA \\
mike.matessa@collins.com}
\and
\IEEEauthorblockN{Siddhartha Bhattacharyya}
\IEEEauthorblockA{\textit{Computer Engineering and Sciences} \\
\textit{Florida Institute of Technology}\\
Melbourne FL, USA \\
sbhattacharyya@fit.edu}
\linebreakand
\IEEEauthorblockN{Randolph M. Jones}
\IEEEauthorblockA{\textit{Applied Cognitive Systems} \\
\textit{Soar Technology}\\
Ann Arbor MI, USA \\
rjones@soartech.com}
\and
\IEEEauthorblockN{Jennifer Davis}
\IEEEauthorblockA{\textit{Applied Research and Technology} \\
\textit{Collins Aerospace}\\
Cedar Rapids, IA U.S.A. \\
jen.davis@collins.com}
\and
\IEEEauthorblockN{Parneet Kaur}
\IEEEauthorblockA{\textit{Computer Engineering and Sciences} \\
\textit{Florida Institute of Technology}\\
Melbourne FL, USA \\
pkaur2022@my.fit.edu}
\linebreakand
\IEEEauthorblockN{Simone Fulvio Rollini}
\IEEEauthorblockA{\textit{Applied Research and Technology} \\
\textit{Collins Aerospace}\\
Rome, Italy \\
simonefulvio.rollini@collins.com }
\and
\IEEEauthorblockN{Natasha Neogi}
\IEEEauthorblockA{\textit{Langley Research Center} \\
\textit{NASA}\\
Hampton VA, USA \\
natasha.a.neogi@nasa.gov }
}

\maketitle

\begin{abstract}
With the rapid advancements in Artificial Intelligence (AI), autonomous agents are increasingly expected to manage complex situations where learning-enabled algorithms are vital. However, the integration of these advanced algorithms poses significant challenges, especially concerning safety and reliability. This research emphasizes the importance of incorporating human-machine collaboration into the systems engineering process to design learning-enabled increasingly autonomous systems (LEIAS). Our proposed LEIAS architecture emphasizes communication representation and pilot preference learning to boost operational safety. Leveraging the Soar cognitive architecture, the system merges symbolic decision logic with numeric decision preferences enhanced through reinforcement learning. A core aspect of this approach is transparency; the LEIAS provides pilots with a comprehensive, interpretable view of the system’s state, encompassing detailed evaluations of sensor reliability, including GPS, IMU, and LIDAR data. This multi-sensor assessment is critical for diagnosing discrepancies and maintaining trust. Additionally, the system learns and adapts to pilot preferences, enabling responsive, context-driven decision-making. Autonomy is incrementally escalated based on necessity, ensuring pilots retain control in standard scenarios and receive assistance only when required. Simulation studies conducted in Microsoft's XPlane simulation environment to validate this architecture’s efficacy, showcasing its performance in managing sensor anomalies and enhancing human-machine collaboration, ultimately advancing safety in complex operational environments.
\end{abstract}

\begin{IEEEkeywords}
Human-Autonomy Interactions, Human-Autonomy Interface, Cognitive Architecture, Adaptive Autonomy
\end{IEEEkeywords}

\section{Introduction}
With the advancements in machine learning, it is expected that autonomous agents/systems will become increasingly intelligent and capable of handling complex emerging situations. This is evident from research explorations evaluating advanced capabilities such as autonomous collision avoidance, advanced air mobility, drone delivery systems, and autonomous cars. Newton et. al \cite{NewtonAIAA2021} discuss the need for integrating new advanced algorithms to ensure safety of passengers and bystanders for autonomous Urban Air Mobility (UAM). They also discuss how to assess vehicle capability to handle contingency scenarios. Gregory et. al \cite{GregoryDDDAS2020} discuss the need for using deterministic and learning models to handle uncertain situations in intelligent systems design and assessment. Although these advancements enhance the expertise of autonomous systems, they make assurance and certification more challenging, particularly when dealing with multiple sensors and varying levels of autonomy.  

In the research effort by Bhattacharyya et. al \cite{BhattacharyyaNFM2018,BhattacharyyaFMAS2021,10131227}, they discuss the design of a framework to include Human-Machine Interfaces and formal methods to verify the correctness of an Increasingly Autonomous System (IAS). While their IAS framework provides valuable insights, it does not include any preference learning or multi-sensor integration. For the assurance of LEIAS designed for contingency management, Neogi et. al \cite{Neogi2021TAITS} investigate chunking, a general learning mechanism in Soar \cite{LairdSoar2012}. However, they do not evaluate human-machine interactions or address scenarios involving multiple sensor failures and autonomous decision-making.

Building upon the Assured Human Machine Interface for Increasingly Autonomous Systems (AHMIIAS) approach \cite{BhattacharyyaFMAS2021,10131227}, we extend the framework to handle multiple sensor inputs and adaptive autonomy levels. In section \ref{sec:RW} related work is discussed, with the proposed methodology in section \ref{sec:RM}. In section \ref{HMILiEAS}, we present enhanced human-machine interactions for multi-sensor scenarios and the extended learning capabilities of the IAS. The design and experiments with the multi-sensor LEIAS are explained in section \ref{sec:LEIAS}. We demonstrate the effectiveness of our approach through comprehensive simulation studies in section \ref{sim}, followed by conclusions in section \ref{sec:conclusion}.

\textbf{Our contribution} is the extension of our previous framework to incorporate:
\begin{itemize}
    \item Enhanced multi-sensor integration (GPS, IMU, and LIDAR) with learning-enabled reliability assessment
    \item Adaptive autonomy levels that respond to sensor reliability and pilot engagement
    \item Extended pilot preference learning across multiple sensor domains
    \item Novel human-machine interface design for transparent communication of learnt behavior to detect, inform and change unreliable sensors
    \item Validation through XPlane simulation studies demonstrating the effectiveness of the enhanced framework in complex scenarios involving sensor anomalies and varying levels of pilot engagement
\end{itemize}

The architecture models, IAS agent, interface design, and simulation results can all be found on our project repository \footnote{https://github.com/loonwerks/AHMIIAS/tree/Year\_3}.

\section{Related work}
\label{sec:RW}
The development of IAS in civil and military aviation has been significantly accelerated by recent technological advancements. Alves et al. \cite{alves2018considerations} discuss the inherent challenges in assuring the safety of IAS, primarily driven by the adaptive and non-deterministic behaviors exhibited by these systems. Traditional authority transfer from autonomous functions to human operators, which has been a cornerstone of safety protocols, is becoming insufficient as autonomy increases. To address these challenges, the authors propose a comprehensive safety assurance framework, particularly for scenarios involving reduced crew operations, where the role of autonomous systems is more pronounced. In the pursuit of enhancing the learning capabilities and adaptability of IAS, several methodologies have been proposed. Mani et al. \cite{ManiAIKE2018} examine a variety of approaches, including deep learning and real-time discovery mechanisms, aimed at improving both system autonomy and the collaborative interface between humans and machines.

One of the critical challenges faced by LEIAS is ensuring safety while operating in dynamic and uncertain environments. Zhu et al. \cite{zhu2021safety} underscore the need for a holistic approach to safety assurance that encompasses both vertical and horizontal integration. Vertical integration refers to the alignment across functional, software, and hardware layers, while horizontal integration focuses on the cohesion across sensing, perception, planning, and control modules. Their work, especially in the context of connected and autonomous vehicles, demonstrates how safety-assured design principles can be effectively applied to multi-sensor systems with varying levels of autonomy.

Research efforts aimed at enhancing adaptive autonomous systems have predominantly concentrated on sensor integration and reliability. In this regard, Model Predictive Control (MPC) has emerged as a promising technique, allowing systems to learn from sensor data and adapt to evolving conditions. Hewing et al. \cite{HewingARCRAS2020} explore different methodologies for learning system dynamics and implementing safe control strategies in environments with sensor uncertainty. In parallel, Learning-Enabled Systems (LESs) have been designed with goal-based, model-driven approaches to ensure self-adaptation and continuous self-assessment \cite{LangfordModalas2021}. The Model-Driven Assurance for Learning-enabled Autonomous Systems (MoDALAS) framework \cite{LangfordModalas2021} addresses the uncertainties that arise in LESs, with a particular focus on sensor reliability and system adaptability in dynamic operational contexts.

The interaction between humans and machines in autonomous systems has also received significant attention. Sadler et al. \cite{sadler2016effects} emphasize the importance of maintaining transparent communication between pilots and autonomous systems, particularly in environments where sensor uncertainties are prevalent. Johnson et al. \cite{johnson2019no} explore adaptive autonomy frameworks that dynamically adjust based on pilot engagement and system state. Meanwhile, Taylor et al. \cite{taylor2003safety} introduce a "Safety Net" system that utilizes flexible autonomy levels to provide cognitive assistance to pilots, helping mitigate the risks associated with spatial disorientation (SD) and loss of situation awareness (SA). However, their approach, being primarily focused on military aviation, has limited applicability to broader autonomous system contexts, as it relies on pre-established plans and pilot inputs, which may not suffice in highly dynamic environments that require real-time decision-making.

The present work addresses these limitations by extending adaptive autonomy frameworks to civil aviation and other safety-critical domains. Specifically, we propose a dynamic, data-driven decision-making process that accounts for sensor reliability and environmental factors, thereby reducing the dependency on human intervention while enabling continuous collaboration between human agents and autonomous systems. Furthermore, our approach incorporates multi-sensor integration to enhance system adaptability across various operational domains, bridging the gap between single-domain and generalized autonomous system applications.

Bhattacharyya et al. \cite{BhattacharyyaNFM2018, BhattacharyyaFMAS2021,10131227} have contributed to the design of human-machine team architectures, focusing on increasing trust and ensuring safety. Their work provides a foundational approach for integrating human factors into autonomous systems, while addressing the system adaptation and safety assurance challenges outlined by Zhu et al. \cite{zhu2021safety}. Recent work by Park et al. \cite{park2020driver} extends this body of research by investigating driver preference learning and adaptive decision-making processes in autonomous vehicular systems. As Herse et al. \cite{herse2018you} and Schneider \& Kummert \cite{schneider2021comparing} have shown, systems that incorporate preference learning can increase trust in the system. Trust can also be increased with system transparency \cite{sadler2016effects}. Building upon these prior works, the present study enhances IAS capabilities by integrating multi-sensor data, pilot preference learning, and adaptive autonomy levels in a transparent manner. We extend the work of Bhattacharyya et al. \cite{BhattacharyyaFMAS2021,10131227} by incorporating dynamic decision-making mechanisms that are grounded in sensor reliability assessment and pilot engagement. Our approach is validated through extensive simulation studies in XPlane, demonstrating its applicability to both aviation and broader safety-critical domains.

\begin{figure}[hbt!]
\includegraphics[ scale = 0.022]{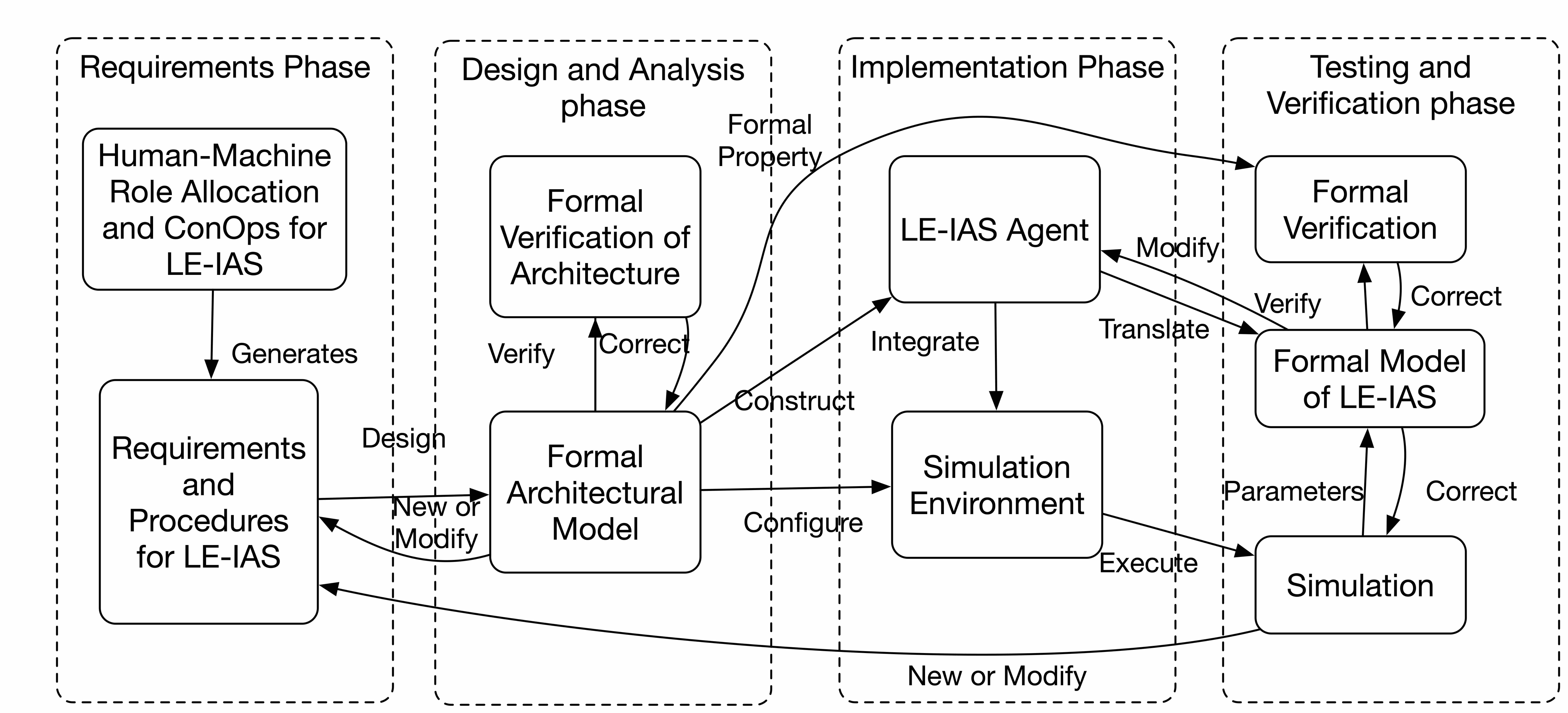}
\caption{Proposed methodology for LEIAS} 
\label{fig:RM}
\end{figure}

\section{Research Methodology} 

\label{sec:RM} 

The methodology adopted for designing LEIAS is depicted in Fig. \ref{fig:RM}. In the Requirements phase, learning scenarios are conceptualized, accompanied by detailed specifications for human-machine interfaces (HMI). These are subsequently aligned with the architectural components necessary for the learning processes. In the Design and Analysis phase, a formal verification of the architecture is conducted to ensure correctness, and the formal architectural model is constructed and verified against formal properties. During the Implementation phase, behavioral specifications are incorporated as learning infrastructure, followed by the design of the learning agent and experimental validation to achieve the intended behavior of LEIAS. The final phase, Testing and Verification, involves a rigorous formal verification process. 

\section{HMI and LEIAS Architecture} 
\label{HMILiEAS} 

\subsection{Human-Machine Interface (HMI)}

Bhattacharyya et al. \cite{BhattacharyyaFMAS2021,10131227} proposed two scenarios to evaluate the shared responsibilities and collaboration between a human pilot and an IAS in Urban Air Mobility (UAM) contexts. The primary scenario, referred to as the Unreliable Sensor Scenario, introduces an anomaly where the GPS sensor's reliability is compromised due to urban canyon effects. The IAS identifies this GPS sensor unreliability and computes the actual position using both Lidar and IMU data, notifying the pilot of the inconsistency. The pilot can either confirm or contest the IAS's interpretation. The scenario is further expanded to integrate learning, enabling the IAS to adapt based on pilot responses and preferences. As part of this enhancement, HMI specifications were refined to convey what the IAS has learned to the pilot. The scenario was extended to include additional sensors, specifically Lidar and IMU. Moreover, the system now has the autonomous capability to switch from an unreliable sensor to an available reliable alternative if the pilot fails to respond within a 5-second window. The agent autonomously performs this sensor switch if the detected error is within a safety threshold.

LEIAS serves as a support system for the pilot by continuously monitoring the operational context, assisting with decision-making processes, and providing monitoring/contingency related information, while ensuring that the pilot retains ultimate authority over final decisions.

\begin{figure} 
\center 
    \includegraphics[width=0.9\linewidth]{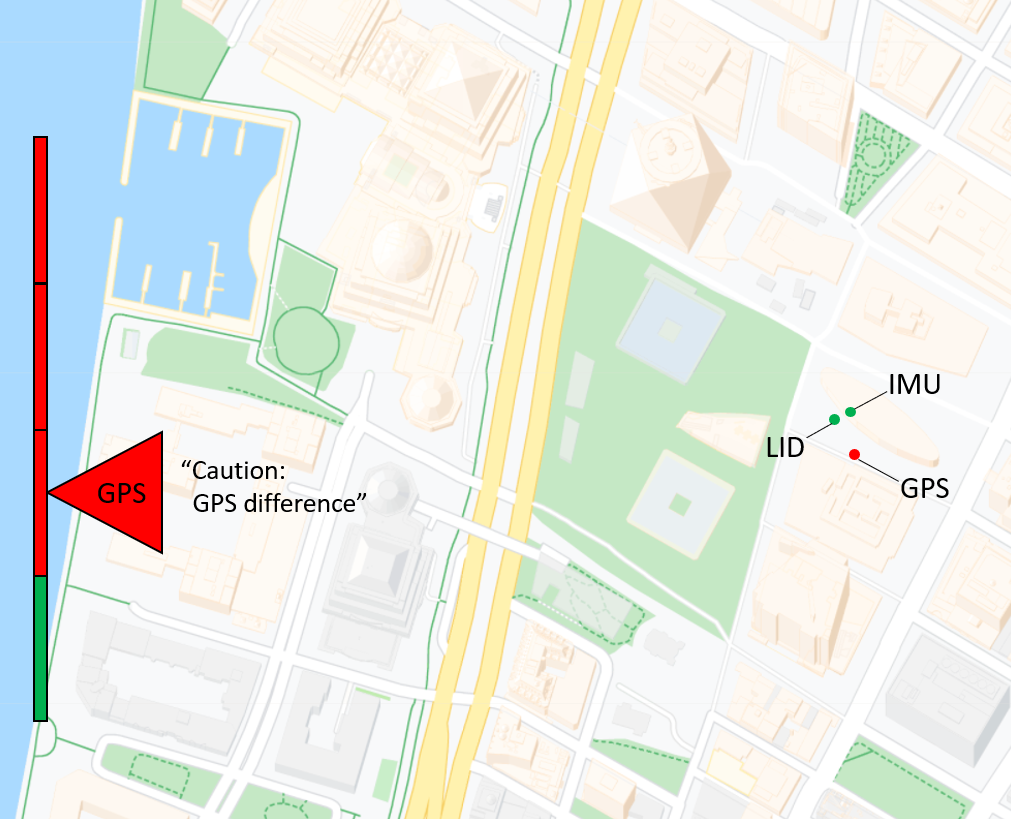} \caption{Learning HMI} 
    \label{learningHMI} 
\end{figure}

The HMI consists of a display with a map, position markers for GPS, Lidar, and IMU sensors, and a warning icon indicating sensor unreliability adjacent to a severity indicator bar (Fig. \ref{learningHMI}). The coloring in the indicator bar gives transparency to the pilot alerting preference learned by the IAS. The position markers illustrate any discrepancies between the sensors' outputs in comprehensible units for the pilot (e.g., within a building, a block, or across town).

The IAS consistently alerts the pilot when an error falls within the safety threshold, refrains from issuing alerts during normal operational ranges, and learns the pilot's preferences for intermediate alert levels (level 1 and level 2). The system indicates alert levels by coloring the corresponding section of the severity bar red (for alerts) or green (for no alerts).

The main objective of these scenarios is to model the interaction and collaborative efforts between the IAS and the pilot. In the Unreliable Sensor Scenario, the IAS identifies the GPS sensor as unreliable by comparing its position data with that obtained from the Lidar and IMU sensors. Errors are represented as abstract values highlighting the pairwise positional discrepancies, without delving into the sensor's internal functioning. The scenario presumes that in this context, the Lidar and IMU sensors offer more reliable data than the GPS sensor, though it is acknowledged that these sensors can also be subject to errors. The scenarios are adaptable for testing various failure modes or different combinations of sensor reliability.

\subsection{Human-Machine Team Architecture Model}
We capture the structure of the model (components and interfaces) in AADL. AADL is a standardized language for embedded, real-time systems. It supports design, analysis, virtual integration, and code generation.
We capture an abstract representation of the behavior of the system and each subcomponent in an assume-guarantee contract in the AGREE. AGREE performs compositional analysis, verifying system requirements based on the composition of the component assume-guarantee contracts. By abstracting the implementation of subsystems and software components into formal contracts, large systems can be verified hierarchically without the need to perform a monolithic analysis of the entire system at once. 

The structure of the human-machine team architecture model builds on the AHMIIAS framework introduced by Bhattacharyya et al.\cite{BhattacharyyaFMAS2021, 10131227}, where two scenarios are distinguished, one characterized by a flawless interaction, and one  where the IAS is subject to a fault or the pilot acts in an unanticipated manner. 
Verification of system requirements in the latter scenario is performed with AMASE \cite{stewart2018safety}, a tool that extends AGREE with safety analysis capabilities, relying on a safety annex that allows to reason about faults and faulty components behaviors. 

\textbf{Nominal behavior.} In the model, at each step two sensors, are respectively marked as ``active'', if currently in use, and ``recommended'', if suggested for the next step, e.g. to replace the active sensor in case of unreliability. 
The IAS interface informs on the active and recommended sensors, and is extended to allow the communication of the discrepancy value, the corresponding range, and an unreliability alert for 
the active sensor; the pilot's interface is extended to provide an acceptance/rejection feedback to an IAS alert. 
Requirements are added in terms of guarantees in the AGREE language, to specify the IAS behavior and to express expectations on the pilot's behavior.
For example, the IAS (i) should not evaluate a sensor as unreliable if the sensor's error range falls into the Normal range. And conversely, (ii) it should determine unreliability if the sensor's error range falls into the \textit{Safety range}. Guarantees are developed that ensure that the IAS correctly evaluates sensor reliability based on their error ranges, maintaining consistency between the Normal range and reliability, and between the Safety range and unreliability for all three sensors in the system.

At the system level, requirements on the interaction between the IAS and the pilot are encoded as AGREE lemmas and subject to formal verification. Whenever a sensor discrepancy falls into the level 1 or level 2 range, the learning process is activated. Broadly speaking, after the pilot responds to an alert in the presence of an active sensor discrepancy, the system's behavior is adjusted based on this feedback. If the pilot's response is \textit{Agree} (acceptance), it should become more "likely" for the IAS to issue an alert for that sensor and discrepancy in the future. Conversely, if the response is \textit{Disagree} (rejection), it should become less "likely" for the IAS to issue such an alert. The precise definition of "likelihood" and the specifics of how it is affected by the pilot's response depend on the particular learning algorithm and implementation; presently, they are not represented at the architecture level.

However, the IAS-pilot communication flow, which enables this learning process, is modeled through guarantees on both the IAS and the pilot's side. For example, the pilot is expected to provide feedback to an alert from the IAS within a given amount of time. Similarly, guarantees are formulated within the AGREE framework to ensure a consistent communication pattern between the IAS and the pilot, providing a framework for the learning process to occur based on the pilot's feedback to sensor unreliability alerts.

\textbf{Unexpected behavior.} 
As part of the safety analysis, an unanticipated behavior is introduced on the pilot's side in the form of a lack of response to an IAS unreliability alert. This is implemented through the \textit{fail active sensor unreliability response} node, which models a scenario where the pilot fails to provide the expected feedback. This node simulates a scenario where the pilot gives no response (\textit{Neutral}) even when an alert is active, contrary to the expected behavior defined in the guarantees.
The rationale behind this fault model is to challenge the system when the assumption that the pilot always gives feedback after being alerted of active sensor unreliability fails to hold. This allows for testing the robustness of the IAS and the overall system in scenarios where expected human-machine interaction patterns are not followed, potentially revealing safety-critical issues or areas for improvement in the system's design. 

Enabling the unanticipated behavior (pilot's lack of response) leads to the violation of two system-level requirements. The first requirement, which encodes the expectation of a pilot's response, trivially does not hold when the response is absent. This requirement should be reconsidered if the original assumptions about the pilot's behavior are modified. The second requirement, defines the conditions under which an unreliable sensor is acceptable as the active sensor. This requirement, states that if an unreliable sensor is the active sensor, one of three conditions must be true. The pilot disagreed with the IAS assessment, the sensor just became unreliable on this timestep, or there was no reliable sensor available on the previous timestep. 
\begin{figure}[hbt!]
\centering
\includegraphics[scale=0.6]{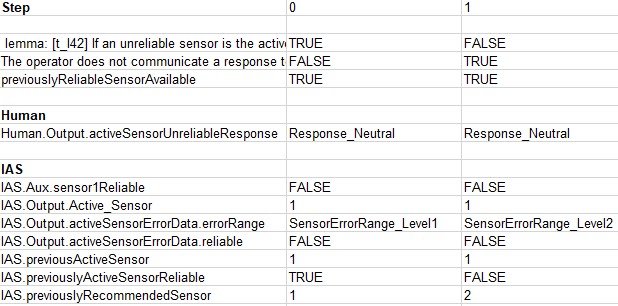}
\caption{Pilot's lack of response - counterexample}
\label{fig:fault_pilot_cex}
\end{figure}

\textbf{Pilot's lack of communication Validation Results.} 
The two-step counterexample generated by AMASE (Fig. \ref{fig:fault_pilot_cex}) corresponds to a situation where the active sensor (\textit{IAS.Output.Active\_Sensor}) is evaluated as unreliable (\textit{not activeSensorErrorData.reliable}) at both steps, but the pilot's lack of feedback means that \textit{Human.Output.activeSensorUnreliableResponse} is neither \textit{Agree} nor \textit{Disagree}. Furthermore, there is a reliable sensor available at the first step, which could have been chosen to replace the active sensor at the second step. This violates the third condition of the lemma, as \textit{previouslyReliableSensorAvailable} would be true. The values at the first step that refer to a previous step are conventionally set, but the violation occurs because the system maintains an unreliable active sensor despite the availability of a reliable alternative, and without the expected disagreement from the pilot. This counterexample highlights the importance of the pilot's feedback in the system's decision-making process and reveals a potential vulnerability when this expected interaction is absent.

The analysis of the Human-Machine Team Architecture Model reveals the importance of robust system design in handling unanticipated human behaviors. With the human and system modeling analysis completed in AGREE, the next step involves representing these interactions in the Soar-XPlane simulation environment. This transition from formal modeling to simulation provides a platform for more dynamic testing and validation of the human-machine interactions.

\section{Implementation of LEIAS}
\label{sec:LEIAS}
The design of a LEIAS in this research effort involves using and evaluating learning capabilities in the cognitive architecture Soar. 

\subsection{Learning in the Soar Cognitive Architecture}

This case study centers on implementing a LEIAS within the Soar cognitive architecture \cite{LairdSoar2012}. Soar is a cognitive architecture designed to replicate human-like cognitive capabilities, providing a robust computational framework for intelligent reasoning and decision-making. It functions by employing knowledge-based rules that allow it to act contextually, simulating cognitive processes such as learning, problem-solving, and decision-making. Soar's versatility supports various learning mechanisms, including chunking, episodic learning, and semantic learning, alongside reinforcement learning \cite{LairdSoar2012}.

Soar's architecture integrates knowledge representation and decision-making processes seamlessly, making it suitable for implementing adaptive learning in complex systems. Its modular nature facilitates the integration of additional data inputs, such as those from multiple sensors, enabling processing and learning. For instance, when working with three distinct sensors—GPS, Lidar, and IMU—Soar can be configured to handle data from each, evaluate their reliability, and adapt based on real-time feedback. This integration supports a dynamic, context-aware decision process, allowing the system to respond appropriately to sensor reliability changes and enhance its performance over time through learning.

Reinforcement learning, a prominent paradigm within machine learning research \cite{SuttonReinforcement2018}, has been adapted in Soar to integrate directly with its decision-making structure \cite{NasonCSR05}. In Soar, the decision-making process involves analyzing the current state, proposing potential actions or inferences (“operators”), evaluating the operators to identify the most suitable option in the present context, and executing the chosen action. This decision cycle operates iteratively and continuously, underlining the intelligent behavior of a Soar agent.

The incorporation of reinforcement learning allows Soar to utilize a reward mechanism to guide behavior based on goal achievement. Reinforcement learning algorithms in Soar assign reward values to previous decisions, enabling the architecture to favor operators that historically led to positive outcomes. After execution, over repeated decision cycles, this results in an adaptive agent that improves its performance by increasingly preferring actions associated with higher rewards.

When applied to the LEIAS context, reinforcement learning can be employed to manage and adapt responses to data from the integrated sensors (GPS, Lidar, IMU). By processing sensor inputs and learning pilot responses, Soar can dynamically adjust its decision-making strategy, such as determining when to switch between sensors in cases of data unreliability. This iterative learning mechanism supports the development of a system that can refine its decision-making capabilities based on real-time feedback, leading to more reliable and contextually accurate behavior in autonomous scenarios.

\subsection{Design of the Integrated Learning Agent}
The focus of this use case is the development of an intelligent IAS that adapts to a human pilot’s preferences regarding alert conditions for potential sensor failures. The emphasis is on learning pilot preferences for three distinct sensors: GPS, IMU, and LIDAR. The LEIAS agent's structure is based on alert regions, as detailed in Section IV B. The design specifies that for normal error ranges, the system will not issue an alert, while for safety-critical error ranges, it will always alert the pilot. For intermediate error levels (levels 1 and 2), the Soar agent adapts to the pilot's preferences by learning for each sensor individually. The specific pilot preferences for alerts at different levels are summarized in Table \ref{tab:pilot-prefs}.

\begin{figure*}
    \centering
    \includegraphics[width=\linewidth]{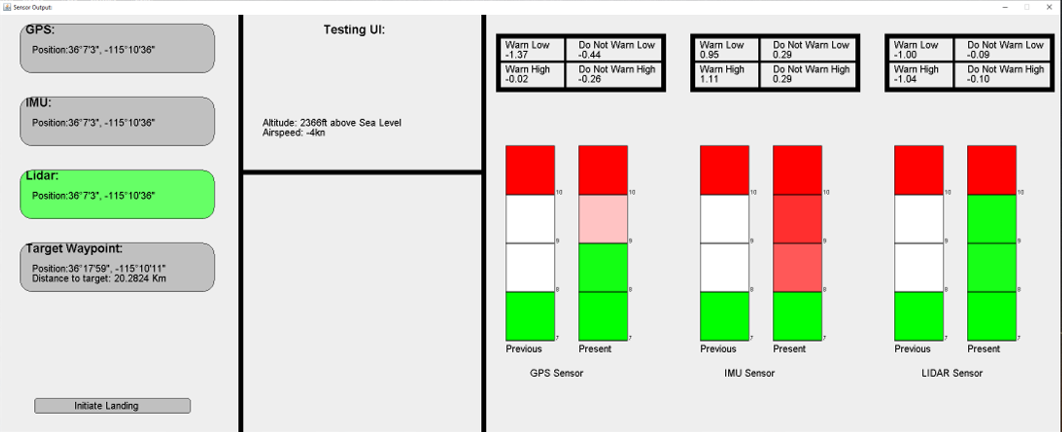}
    \caption{Learning aware HMI}
    \label{fig:viz}
\end{figure*}
\begin{table}[htbp!]
  \caption{Pilot warning levels preference for different sensors}
  \label{tab:pilot-prefs}
  \centering
    \begin{tabular}{ccc}
    \toprule
 \textbf{Sensor} & \textbf{Level 1} & \textbf{Level 2} \\
 \midrule
    \textbf{GPS}    & Warn         & Do Not Warn   \\
    \textbf{LIDAR}  & Warn         & Warn          \\
    \textbf{IMU}    & Do Not Warn  & Do Not Warn   \\
 \bottomrule
\end{tabular}
\end{table}

The learning mechanism is reinforced over time through a feedback system in which the agent receives a positive reward when the pilot acknowledges and accepts an alert, and a negative signal (punishment) when the pilot rejects an alert. The reinforcement learning algorithm used by the Soar agent allocates portions of these reward or punishment signals to the agent’s decisions, enabling adaptive behavior refinement. Each new interaction enhances the agent’s decision-making, allowing it to align more closely with pilot preferences and improve future alerting accuracy.

This continuous learning loop ensures that the LEIAS agent becomes more adept at determining the conditions under which alerts should be issued or withheld, based on observed pilot feedback. By integrating sensor data and adapting through reinforcement learning, the system is capable of providing a more personalized and efficient human-machine interaction tailored to pilot-specific preferences across different sensors. Furthermore, the HMI design provides visual display of what has been learnt by the agent to the pilot.

\subsection{Experiments for LEIAS} To conduct the experiments, the learning architecture was designed to integrate the learning infrastructure, environmental interactions, and the symbolic decision logic-based learning agent. Training and testing cycles were performed to validate the learning process.

The \textbf{learning infrastructure} encompassed all the necessary components and interactions required to facilitate data generation and management. Key infrastructural requirements included data storage to support the multiple cycles needed for reinforcement learning. Additionally, thresholds for alerting levels and pilot response data needed to be stored. To streamline data collection, the pilot's responses were scripted within the infrastructure.

\subsection{Experimental Setup and Results for LEIAS} 
The learning process was organized into distinct phases comprising learning trials and testing trials. In each learning trial, random errors were introduced, prompting the IAS to decide whether to alert the pilot. If the decision aligned with the preferences of the scripted pilot, the agent was rewarded with a score of +1. Conversely, if the decision diverged from the pilot's preferences, a penalty of -1 was applied. In this experimental setup, the scripted pilot expected the IAS to issue a warning whenever the error reached or exceeded level 2 (error value $\geq$ 9). During testing trials, the introduced sensor error incrementally increased until it reached the safety threshold, providing validation for the agent's learned behavior.

Various rule selection policies were explored, including Boltzmann distribution with high and low temperatures and simulated annealing. Simulated annealing involves a gradual learning process starting with a high learning rate or temperature, which decreases progressively. The higher initial temperature allows for exploration of a broader set of potential solutions. In Boltzmann learning, a high acceptance rate can facilitate accepting suboptimal moves initially. As the temperature or acceptance rate declines, the focus shifts toward refining and optimizing the solution. The choice of cooling schedule and temperature is critical to maintaining stability and responsiveness; improper tuning can result in either convergence to suboptimal solutions or extended convergence times.

Fig. \ref{fig:viz} displays the real-time decision-making process and the pilot alerting preferences as learned by the LEIAS agent for each sensor (GPS, IMU, and LIDAR). In this visualization, the left panel shows the current positional data for each sensor and the target waypoint, alongside controls such as the “Initiate Landing” button. The center panel displays current altitude and airspeed. The right panel provides detailed insights into the learned alerting behavior for the GPS, IMU, and LIDAR sensors.

Each vertical bar represents alert levels (green for no alert, white for level 1 alert, and red for level 2 alert), showing both past and present conditions. The numerical values above the bars denote the reinforcement learning scores for each decision criterion. Specifically:
\begin{itemize}
    \item GPS Sensor: The learned model indicates a preference for issuing alerts at lower error thresholds (indicated by the red section at level 9), aligning with the pilot’s preferences for warnings in potentially critical situations.
    \item IMU Sensor: The IMU sensor's results illustrate a more balanced distribution of warnings and non-warnings, with the presence of alerts primarily at higher error levels.
    \item LIDAR Sensor: The LIDAR data display shows consistent behavior in alerting when errors are significant, indicating that the learning agent has adapted to provide appropriate warnings in level 1 and level 2 conditions.
\end{itemize}

The scores at the top of the UI (e.g., "Warn Low", "Do Not Warn Low", "Warn High", "Do Not Warn High") reflect the reward-based adjustments made by the agent. These scores illustrate the agent’s confidence in making decisions based on pilot preferences:
\begin{itemize}
    \item Positive scores (e.g., 0.95 for “Warn Low” in IMU) suggest that the system is confident in its decision to alert the pilot at lower levels of error.
    \item Negative scores (e.g., -1.00 for “Warn Low” in LIDAR) indicate areas where the system has learned that alerting may not align with pilot expectations or preferences.
\end{itemize}
\begin{figure}[h]
    \centering
    \includegraphics[width=1.1\linewidth]{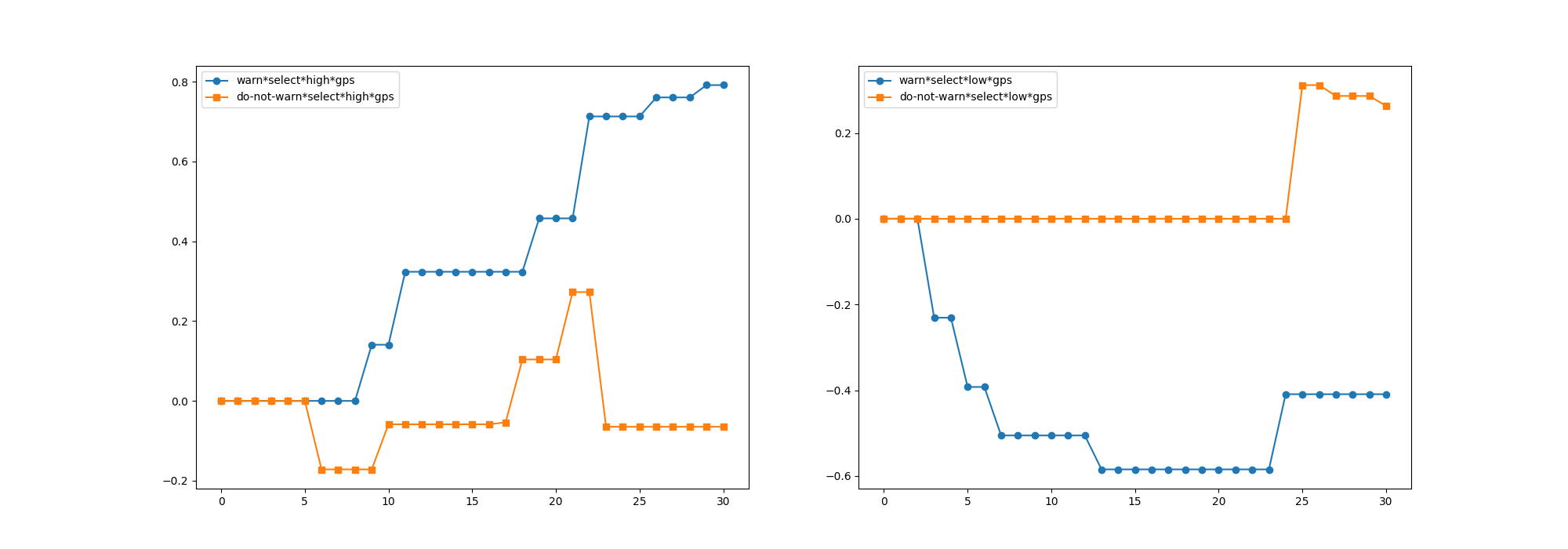}
    \caption{The decision-making process of the LEIAS agent for the GPS sensor, showing the comparison between warning  (in blue) and non-warning (in orange) actions. The reinforcement learning scores reflect the alignment of these decisions with the pilot's alerting preferences.}
    \label{fig:gps}
\end{figure}
\begin{figure}[h]
    \centering
    \includegraphics[width=1.1\linewidth]{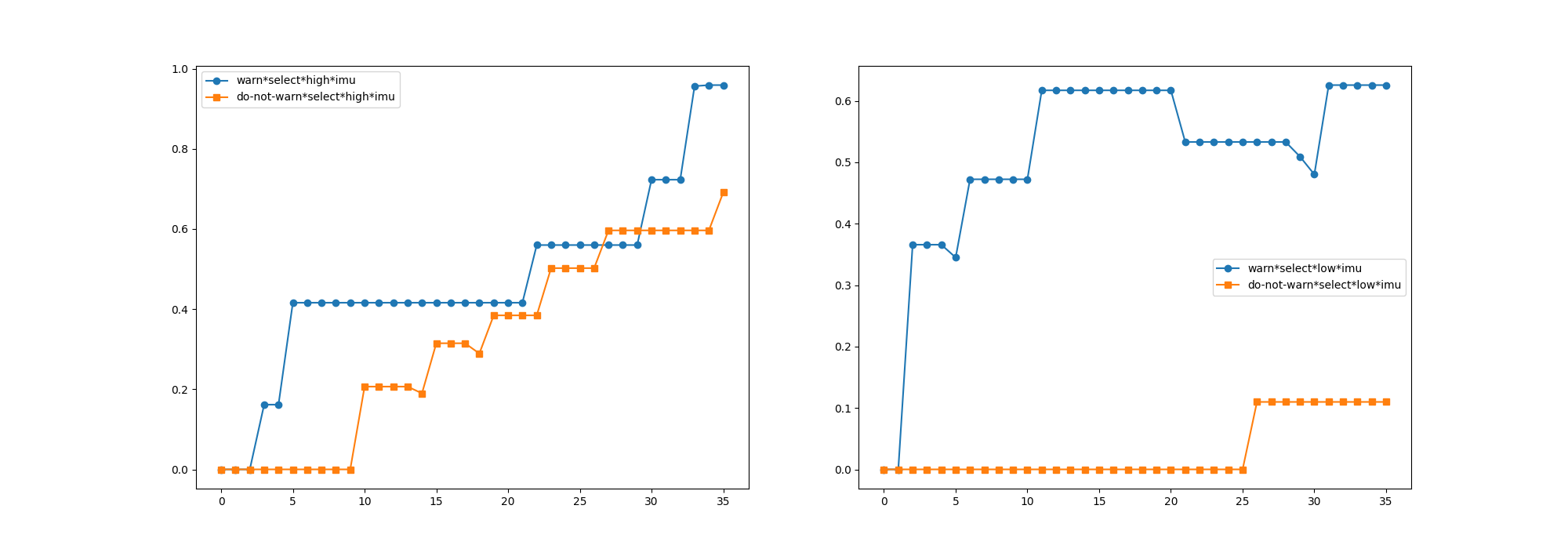}
    \caption{The IMU sensor chart presents the agent’s learned behavior, showing the comparison between warning (in blue) and non-warning (in orange) actions. The score trends indicate how the agent adapted to higher and lower error levels based on reinforcement learning feedback.}
    \label{fig:imu}
\end{figure}
\begin{figure}[h]
    \centering
    \includegraphics[width=1.1\linewidth]{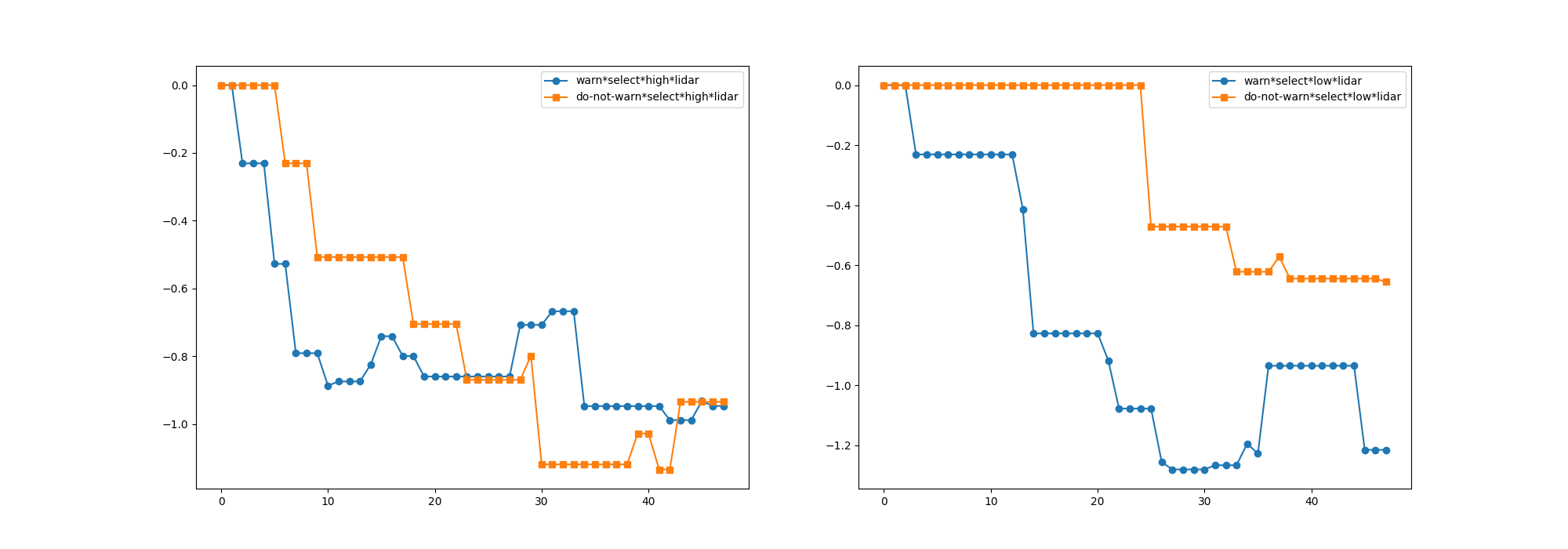}
    \caption{The alerting behavior of the LEIAS agent for the LIDAR sensor, highlighting consistent warning (in blue) and non-warning (in orange) decisions.}
    \label{fig:lidar}
\end{figure}

The visualizations depict the learned alerting behavior of the LEIAS agent for three key sensors: GPS, IMU, and LIDAR. Each plot (Fig \ref{fig:gps}, \ref{fig:imu}, \ref{fig:lidar}) shows the agent's decision-making process and confidence scores in responding to different error thresholds. The GPS sensor chart reveals a stronger inclination to issue warnings at lower error levels, indicating alignment with the pilot’s preference for early alerts. The IMU sensor visualization displays a more balanced response, where warnings are more frequent at higher error levels, demonstrating the agent's adaptation to pilot preferences through learning. The LIDAR sensor results highlight a consistent trend in issuing warnings, especially at significant error thresholds, suggesting the agent's capability to handle critical warning levels effectively. These visualizations underscore the agent's ability to learn and apply reinforcement-based strategies to match pilot-specific alerting expectations.

Overall, these results demonstrate that the LEIAS agent successfully learned and adapted to pilot-specific alerting preferences through reinforcement learning, adjusting its decision-making behavior based on trial feedback. The testing scenarios verified that the agent could autonomously respond to sensor data, issuing warnings or refraining from doing so in alignment with learned pilot behavior.
\section{Simulation Studies and Discussion} 
\label{sim}

The simulation was carried out using the XPlane simulation environment, encompassing all three learning scenarios. Once the agent had learned the threshold values for each of the three sensors, it was tested through simulations to evaluate its performance under autonomous conditions. The trials included scenarios where the agent operated independently without pilot input, demonstrating its ability to autonomously switch sensors when necessary. These simulations provided insights into how the agent adapted to sensor errors and validated its learned behavior in real-time, ensuring that the system met the intended alerting and decision-making criteria.
\section{Conclusion and Future work}
\label{sec:conclusion}
The implementation and experimentation of the LEIAS presented in this study have demonstrated the feasibility of incorporating reinforcement learning within the Soar cognitive architecture to enhance human-machine interaction in complex aviation environments. By focusing on pilot preferences for sensor-based alerts, the LEIAS agent successfully learned and adapted its decision-making process, ensuring timely and accurate alerts aligned with pilot expectations. The integration of GPS, IMU, and LIDAR sensor data provided a comprehensive framework for testing and validating the learning capabilities of the system, confirming its ability to autonomously respond to sensor reliability changes and pilot feedback. The results from the experimental trials highlighted the effectiveness of using reinforcement learning for adaptive behavior in decision-making processes. Simulations validated that the LEIAS agent could accurately discern when to issue warnings based on varying error levels and pilot responses. 

Future work should focus on extending the current framework to integrate real-world sensor data instead of simulated data, which would enhance the system's reliability and practical applicability by addressing additional complexities such as noise and varying environmental conditions. Additionally, research could further explore enabling the LEIAS agent to adapt to different pilot profiles and preferences, creating a flexible learning mechanism that adjusts to diverse pilot behaviors and requirements. Moreover, enhancing the Human-Machine Interface (HMI) design to offer more intuitive visualizations and interactive modes for pilots would bridge the gap between human understanding and autonomous system behavior, making the interface more user-friendly and informative for real-time decision-making. Investigating additional safety measures, including redundancy checks and fault tolerance mechanisms, would contribute to the robustness of the LEIAS agent, ensuring consistent performance under different operational stresses.

\bibliographystyle{unsrt}
\bibliography{refs}
\end{document}